\documentclass[11pt,a4paper,titlepage,superscriptaddress,nofootinbib]{revtex4-1}
\usepackage{amsfonts,amscd,amsmath, amssymb,graphicx,color}

\usepackage{amssymb}

\begin{document}


\title{St\"uckelberg  Formulation of Holography}

\author{Gia Dvali}
\affiliation{Arnold Sommerfeld Center, Ludwig-Maximilians-University, Theresienstr. 37, 80333 M\"unchen, Germany}
\affiliation{Max-Planck-Institut f\"ur Physik, F\"ohringer Ring 6, 80805 M\"unchen, Germany}	
\affiliation{Center for Cosmology and Particle Physics, Department of Physics, New York University, 4 Washington Place, New York, NY 10003, USA}
\author{Cesar Gomez}
\affiliation{Arnold Sommerfeld Center, Ludwig-Maximilians-University, Theresienstr. 37, 80333 M\"unchen, Germany}
\affiliation{Instituto de F\'{\i}sica Te\'orica UAM-CSIC, \\Universidad Aut\'onoma de Madrid, Cantoblanco, 28049 Madrid, Spain}
\author{Nico Wintergerst}
\affiliation{The Oskar Klein Centre for Cosmoparticle Physics, Department of Physics, Stockholm University, AlbaNova, 106 91 Stockholm, Sweden}

	\begin{abstract}

   We suggest that holography can be formulated in terms of the 
 information capacity of the St\"uckelberg degrees of freedom 
that maintain gauge invariance of the theory in the presence of an information  boundary.  These  St\"uckelbergs act as qubits that account for a certain  fraction of quantum information. 
Their information capacity is measured by the ratio of the inverse St\"uckelberg energy gap to the size of the system.  Systems with the smallest gap are maximally holographic.
 For massless gauge systems this information measure  is universally equal to the inverse coupling evaluated at the systems' length scale.   In this language it becomes very transparent  why the St\"uckelberg information capacity of black holes saturates the Bekenstein bound and accounts for the entire information of the system.  
The physical reason is that the strength of quantum interaction is bounded from below by the gravitational coupling, which scales as area.  Observing the striking similarity between the  scalings of the energy gap  
of the boundary St\"uckelberg modes and the Bogoliubov modes of critical many-body systems, we establish a connection between holography and quantum criticality through the correspondence between  these modes.

	\end{abstract}
	\maketitle

\section{St\"uckelberg Qubits.}   

   The scaling of Bekenstein entropy \cite{Bekenstein} as the black hole area in Planck units, $S_{BH}  = (R/L_P)^2$, suggests that the
   information is stored in some boundary qubit degrees of freedom, one per each Planck-area-size pixel. 
   This approach is usually called the principle of black hole holography \cite{tHooft}.     
  Without committing to the literal validity of such a picture, it is nevertheless necessary to understand the fundamental reason behind the area law.   One microscopic  explanation \cite{us,critical} (see, \cite{gold, hair, portrait} for further discussions) is that the area law is a reflection of the scaling of the quantum gravitational coupling, $\alpha(R)$, of gravitons of wavelength $R$.  Notice, that the most basic property of the inverse gravitational coupling  is to ``run" with the length-scale $R$ as  the {\it area}, 
  \begin{equation}
  {1 \over \alpha (R)}  \, = \,  {R^2 \over L_P^2} \,. 
  \label{gravitycoupling}
  \end{equation} 
Moreover note that the above relation between the quantum  gravitational coupling and the area holds in an arbitrary number of dimensions. 
 Thus, in the picture of \cite{us,critical}, the black hole entropy is fundamentally defined  by  
 the inverse graviton coupling
  \begin{equation}
   S_{BH}   \, = \,   {1 \over \alpha (R)}  \,,  
  \label{entropy}
  \end{equation} 
 and the area law is a direct consequence of  (\ref{gravitycoupling}).  The microscopic reason behind 
 (\ref{entropy}) according to \cite{critical, gold} is the quantum criticality of the black hole state, which 
 is characterized by the appearance of $N \, = \, \alpha^{-1}(R)$ collective  Bogoliubov-Goldstone modes with energy gap  $\Delta  =  \alpha(R) { \hbar \over R}$. The corresponding entropy  is 
 obviously given by (\ref{entropy}). 
 
   Hence, in this picture holography is an emergent phenomenon that results from the 
   {\it quantum criticality}  of the underlying gravitational system, for which both the number $N$ as well as the energy gap  $\Delta$ of the qubits is determined by the gravitational coupling (\ref{gravitycoupling}).  As we shall see, we shall discover a connection, strikingly similar to (\ref{entropy}),   between the information capacity and the inverse coupling of the system, from a completely different  approach.
  This allows us to draw an interesting conclusion about the nature of the information qubits  of holographic systems, in which  gauge-redundancy and quantum criticality appear as two sides 
  of the same coin.

 In order to formulate this alternative approach, in the present paper we would like to ask whether there exists a simple way for understanding the area law from a fundamental gauge principle, without the knowledge of the microscopic picture of the black hole interior.   

 We shall be interested in systems that contain sub-systems separated by surfaces that act 
 as boundaries from the information point of view. 
Consider two regions of the space, $A$ and $B$, separated by a closed surface $S$.  Let us assume that due to some dynamical reason the retrieval of information from the region $B$ into $A$ is highly suppressed or simply impossible. We shall say that for an observer in the region $A$ (we shall call her Alice) the surface $S$ acts as an {\it information boundary}. 
 The suppression of the information-retrieval probability can be, for example, due to an energy barrier that prevents some particle species crossing  over from
 $B$ to $A$ (see examples below).  One natural example that fits within our definition of the  information boundary is a black hole horizon. In this case the region $B$ is identified with the interior of the black hole and Alice is an external observer. 
   
 The general question is:  How can Alice characterize the information-storage capacity of the hidden region $B$? 
 One known possibility is that Alice traces over the degrees of freedom in  $B$ and ends up with a density 
 matrix. Naturally, in such a treatment  an entanglement  
 entropy  is attributed to the region $B$.  The problem with this information measure is that it is based on a complete ignorance about the region 
 $B$.  Consequently, the resulting entanglement entropy, instead of giving a real information measure, gives only an upper bound on the information-storage capacity of the region $B$. In reality,  the true information-capacity may be much smaller.  For black holes, the entanglement 
 entropy gives the same scaling with area as the Bekenstein entropy.  However, this fact does not constitute a satisfactory microscopic explanation of the black hole entropy, but rather a reflection of a simple  coincidence, namely that a  black hole happens to saturate the Bekenstein entropy bound.  Correspondingly, the two entropies agree in scaling.   
   Therefore,  in our approach we shall not rely on entanglement entropy. 
 
 The question we would like to ask  is whether there is a more efficient measure of information-storage capacity that Alice could derive without the detailed knowledge of the microscopic properties of the system $B$.  
 
  In the present paper we would like to point out that in systems in which the information carriers are some sort of gauge degrees of freedom,  the information-storage capacity of the interior region can be understood in terms of the energy gap of the boundary 
St\"uckelberg degrees of freedom that are necessary for maintaining gauge redundancy of Alice's description in the presence of the information boundary $S$.

 In order to outline the general argument, let us assume that Alice deals with a bulk gauge theory of a gauge field 
 $h$ with a Lagrangian density $L(h)$, that under gauge transformation shifts by a total derivative $\delta L = d\Omega$. 
   In the presence of the information boundary $S$, this generates a surface term  $\int_{S} \Omega$, where 
  the integral is taken over the world-volume of the boundary surface $S$.  
  By gauge redundancy,  such a term must be cancelled by a shift of some St\"uckelberg field, $\eta$. 
   
  Thus,  the effective theory 
 (from the information perspective) that Alice deals with is a bulk gauge theory plus  the world-volume action of surface $S$ with the St\"uckelberg degrees of freedom, 
 \begin{equation}
{\mathcal  S} \, = \,   \int_{A} L(h)   +  \int_{S} \eta \,, 
 \label{action}
 \end{equation}   
 such that  the variation of $\eta$ is $\delta \eta = - \Omega$. 
 
  It is important to stress here what is the key point of the former argument. Alice needs to have in her region  {\it manifest} gauge invariance.  The necessity of imposing this condition on her physics forces her to decorate what for her is acting as an information boundary with new physical degrees of freedom. In other words, she is deriving part of the actual dynamics of the information boundary by imposing a manifest gauge invariance in the region that is, in information terms, accessible to her.

  The new degree of freedom $\eta$ provides the natural qubits that store (part of the) information about the region $B$.  The elementary qubits can be labelled either by the world-volume coordinates of the boundary, or equivalently by the world-volume momentum modes up to some maximal momentum.   For example, for a spherical boundary these can be labelled by the usual spherical harmonics $l,m$.     
   Obviously, the total number of elementary qubits scales as the {\it area}  of the boundary $S$.  Thus, the quantum information can be stored in the occupation numbers of different world-volume momentum modes.  For example, in the spherical case the quantum information can be included in a  set of occupation numbers $n_{m,l}$ labeled by the 
 spherical harmonics $m,l$.

  The capacity of information-storage is measured by  the energy gap of the qubit $\eta$.  
   Let this gap be $\Delta$. Let us also assume that the size of the region $B$ is  $R$. 
   Then, a useful measure of the information-storage capacity is the following quantity, 
   \begin{equation}
    {\mathcal  I}  \equiv  {\hbar \over R\Delta} \, .
     \label{Inform}
   \end{equation}
   The physical meaning  of  the quantity $ {\mathcal  I} $ is the following. In a generic system of size $R$, with weakly interacting degrees of freedom,  we can store a single bit of information at the energy expense $\sim \hbar /R$.  Indeed, 
  this energy roughly measures the spacing between the ground-state and the first excited energy-level in such systems. 
Thus, the quantity $ {\mathcal  I} $ gives us a measure of how costly is, relative to this energy, 
the storage of a single bit of information in the  St\"uckelberg degree of freedom $\eta$. 
 If $ {\mathcal  I}  \gg 1$ the system is an efficient storer of information. In the opposite case, $ {\mathcal  I}  \ll 1$, the information-storage capacity of the system is very poor \footnote{There exist a different natural interpretation of ${\mathcal  I}$ that Alice may immediately think about. Since Alice knows the size $R$ of the region $B$, she can infer the quantum uncertainty in the energy stored in this region as $\frac{\hbar}{R}$. However, for her this uncertainty is {\it resolved} into ${\mathcal  I}$ St\"uckelberg boundary states per each qubit given by $\eta$. Hence, she can already infer that ${\mathcal  I}$ measures a part of the {\it degeneracy} of the ground state, whatever may be the underlying reason from the microscopic physics point of view, describing the region-$B$.}. 
 
   Notice that the quantity ${\mathcal I}$ measures the part of the entropy that is stored in the 
   St\"uckelberg modes.   This is simple to understand. 
  In order to evaluate this part of the entropy, we have to compute the number of 
 St\"uckelberg  states that populate the energy levels within the energy gap $\hbar /R$ and then take the log of this number.  Restricting ourselves by considering
  the first excited level for each qubit, the state with $N$ excited qubits has energy 
  $E = N \Delta$.  Restricting this energy to be $ E < \hbar /R$, we easily get that the maximal 
allowed number of the excited qubits is  $N_{max}={\mathcal I}$.   The corresponding number of states 
obviously is $\sim 2^{\mathcal I}$ and the resulting contribution to the  entropy is ${\mathcal I}$. 

We can derive a more general expression relating the entropy to the number of St\"uckelberg modes $N$ and the maximal number of excited modes ${\cal I}$. Through simple combinatorics, we obtain
\begin{equation}
\label{eq:entropy}
e^S = \sum_n^{\cal I}{N + n - 1 \choose n} = \frac{\Gamma(1 + N + {\cal I})}{\Gamma(1 + {\cal I})\Gamma(1 + N)}\,,
\end{equation}
where the sum is to count all states with occupation $n \leq {\cal I}$.
For large $N$ and ${\cal I}$, the leading order expression for general $N$ and $\Delta$ is 
\begin{equation}
\label{eq:entropy_large}
S \approx {\cal I}\left(1 + \log\frac{N}{{\cal I}}\right)
\end{equation}
   
    How big is $\Delta$?  On very general physical grounds, we can estimate it in the following way. 
   Let $m$ be the characteristic mass scale of the theory in the boundary region.  For example, 
 if the mass gap is generated by fluctuations of some massive particles composing the boundary, 
 the scale $m$ will be set by the mass of these degrees of freedom.  In case of no pre-existing mass gap in the theory, the natural scale will be given by the inverse size of the system, 
 $m= \hbar/R$.   Let $\alpha$ be the quantum coupling of the fluctuating degrees of freedom. 
 Then the expected energy gap for the St\"uckelberg modes is  
  \begin{equation}
      \Delta_\text{St\"uckelberg} = \alpha(m) \,  m \, ,  
   \label{gap!}   
  \end{equation}   
  where  $\alpha(m)$ is the coupling evaluated at the energy scale $m$ (or equivalently 
  at the length-scale $\hbar/m$).     
  Thus, the information capacity is, 
   \begin{equation}
      {\mathcal  I}  \equiv  {1 \over \alpha (m)} \,  {\hbar \over Rm} \, .
     \label{Im}
   \end{equation}
   
  On the other hand, if the gauge theory in question is massless, then we must take $m= \hbar/R$ and 
  the St\"uckelberg qubit energy gap becomes, 
   \begin{equation}
 \Delta_\text{St\"uckelberg} \, = \, \alpha(R) {\hbar \over R}  \, .
 \label{gapmzero}
 \end{equation} 
 The information measure is therefore given by,  
    \begin{equation}
      {\mathcal  I} \, = \,   {1 \over \alpha (R)} \, .
     \label{Imzero}
   \end{equation}
  Thus, for a gapless gauge theory that can dynamically produce the information boundary, the 
  information capacity is measured by the inverse quantum coupling, exactly in the same way 
  as this was happening, according to (\ref{entropy}), for Bekenstein entropy! 
  In other words,  applying (\ref{Imzero}) to the black hole horizon, and taking into account that the gravitational coupling at the scale $R$ is given by (\ref{gravitycoupling}), yields
        \begin{equation}
      {\mathcal  I}_{BH}  \, = \, {R^2 \over L_P^2} \,
   \end{equation}     
    For a black hole, the number of St\"uckelberg modes $N$ and the information measure ${\cal I}$ coincide. Hence we recover from Eq.\eqref{eq:entropy_large} the well-known area law,
    \begin{equation}
    S =   {R^2 \over L_P^2} =  S_{BH}  \, .
    \label{areaI}
    \end{equation}
    
    Let us summarize our main message.  In any gauge theory, Alice can characterize the part of the quantum information hidden in the region $B$ in terms of the St\"uckelberg qubits. 
    For an arbitrary gapless gauge system,  this information is measured by 
   (\ref{Imzero}). But, for a generic system, it accounts only for a part of the hidden information. 
    For black holes, due to the scaling of the gravitational quantum coupling $\alpha(R)$, the 
  amount of information that can be accounted by   St\"uckelberg qubits  saturates the bound and thus accounts for the entire information of the hidden region.  
  
  Finally, an interesting connection emerges between the St\"uckelberg formulation presented here   and the previous picture of holography developed in \cite{us,critical}.  As explained above, in the latter picture, 
 holography results from the many-body description of the black hole interior in terms of a
 critical graviton condensate delivering Bogoliubov modes with tiny energy gaps. In the present paper 
 we understand holography in the gauge-redundancy language of an external observer in terms of a
 very similar energy gap of the St\"uckelberg modes. This coincidence suggests that we are dealing with two languages describing the same physics.  
   The striking similarity between (\ref{Imzero}) and (\ref{entropy}) indicates a  
 possible fundamental connection between holography and quantum criticality. This connection  maps the St\"uckelberg qubits of Alice's description onto the  Bogoliubov modes of the many-body quantum critical
 state of \cite{us,critical}.   
    
    In the rest of the paper we  shall elaborate on these ideas.

        \section{Spin-2 Case} 
    
     We shall now discuss our idea in more details. 
   The first part of our argument is generic for any bounded system with gauge redundancy. 
   Let us therefore illustrate it for linearized gravity. The non-linearities can be taken into account, but they do not add anything to the essence of the phenomenon.  
      
 Consider the action of linearized Einstein gravity,  
   \begin{equation}
 S_E \, = \,  \int d^4x L(h) \,  =  \, \int d^4x h^{\mu\nu} {\mathcal E} h_{\mu\nu} \,,  
 \label{actionE}
 \end{equation} 
  where ${\mathcal E} h_{\mu\nu} \, \equiv \,  \Box h_{\mu\nu} \, - \, \eta_{\mu\nu} \Box h \, - \, \partial_{\mu} \partial^{\alpha} h_{\alpha\nu} \, - \, \partial_{\nu} \partial^{\alpha} h_{\alpha\mu}  \, + \, \partial_{\mu} \partial_{\nu} h \, + \, 
  \eta_{\mu\nu} \partial^{\alpha}\partial^{\beta} h_{\alpha\beta}$
 is the linearized Einstein tensor and $h\equiv h_{\alpha}^{\alpha}$.   This system exhibits a gauge redundancy under  the following transformation,  
\begin{equation} 
h_{\mu\nu} \rightarrow h_{\mu\nu} + \partial_{\mu} \xi_{\nu}
+ \partial_{\nu} \xi_{\mu}\,,
\label{gauge} 
\end{equation}
where $\xi_{\mu}(x)$ is a transformation parameter vector. 
   Under this transformation, the Lagrangian density shifts by a total derivative, 
 \begin{equation} 
 \delta L \,  = \,  \partial_{\mu} \Omega^{\mu}\, ~~ {\rm where}~~ \Omega_{\mu} \equiv 2\xi^{\nu} {\mathcal E} h_{\mu\nu} \, .  
 \label{changeL}
 \end{equation}
Let us now introduce a boundary.  The role of it  can be played by an arbitrary closed two-surface described by the target space coordinates 
$X^{\mu},~\mu=0,1,2,3$ and the world-volume coordinates $y^{a}, ~a=0,1,2$.   At the moment, the precise origin of the
boundary is not important for us. For instance, it can represent a dynamical solution of the theory, such as, e.g., a black hole horizon or a brane bubble. Alternatively, it  can be imposed by hand as a mathematical surface.  In either case, what is important for us is that the boundary separates the system into two sub-systems.  We now wish to describe the gauge redundancy from the point of view of the exterior sub-system. 

  Alice sees a gauge theory on a space with boundary.  In this description,  
 naively,  under the gauge shift (\ref{gauge}) the action changes by a boundary term
 \begin{equation} 
 \delta S \, = \int dX^{\mu}\wedge dX^{\nu}\wedge dX^{\alpha} \epsilon_{\mu\nu\alpha\beta} \Omega^{\beta} \, .
 \label{varaction}
 \end{equation} 
   However, since we started by a manifestly gauge-redundant theory, the same redundancy must hold 
   in the correct description of the sub-system. 
  In particular, this is obvious in cases in which the boundary represents a solution of the full gauge-invariant theory.  
   
   The only way to accommodate gauge invariance is  to admit for the boundary to host a new degree of freedom, $\eta_{\mu}$,  that acts as a St\"uckelberg field for maintaining the original gauge invariance.  The action of this new degree of freedom is fixed from the above condition and has the following form, 
   \begin{equation} 
 S_{\eta} \, = \int dX^{\mu}\wedge dX^{\nu}\wedge dX^{\alpha} \epsilon_{\mu\nu\alpha\beta} \eta^{\beta} \, . 
 \label{actionstuck}
 \end{equation} 
  Obviously, the combined action,
    \begin{equation} 
 S =  S_E + S_{\eta} \,,  
 \label{totalaction}
 \end{equation} 
 is invariant under (\ref{gauge}) provided $\eta_{\mu}$ shifts as 
    \begin{equation} 
 \eta_{\mu}  \rightarrow   \eta_{\mu} -  \Omega_{\mu} \,,  
 \label{shifteta}
 \end{equation} 
  with $\Omega_{\mu}$ given by (\ref{changeL}).  
 Notice that $\eta_{\mu}$ is a fully legitimate degree of freedom that transforms as a scalar from the point of view 
of the boundary world-volume theory.  The very existence of this St\"uckelberg degree of freedom follows solely 
from the requirement of symmetry under small diffeomorphisms. However, what promotes them into the physical carriers 
of information is the interaction.

Classically these modes are exactly gapless and can be labeled by  world-volume coordinates of the boundary theory.  
 This means that the number of St\"uckelberg qubits scales as area of the boundary!
This certainly rings a bell,  and makes it tempting  to identify  the  St\"uckelberg fields as the holographic degrees of freedom 
for a black hole.   However, the same coincidence also raises the question: Why are not 
all information boundaries holographic? We shall address this question in the rest of the paper.

 \section{Quantum Generation of Energy Gap and Information Counting} 
 
 In our derivation of boundary St\"uckelberg degrees of freedom, we have never used black hole properties.   Our argument solely relied on gauge invariance and is therefore generic for an arbitrary boundary.  So, what  is special about the black hole horizon? 
   As we shall explain now, the question is closely related to the question of generation of an energy gap in 
  boundary St\"uckelberg modes.  As discussed above, this gap  defines the information-capacity (\ref{Inform}) of the St\"uckelberg modes, which in general can account for a certain fraction of the 
  total information capacity of the bounded inner system.    Black holes turn out to be the systems with the smallest  St\"uckelberg energy gap, which is just enough to saturate the
bound on information. With such a small gap,  black hole horizon St\"uckelbergs can account for the entire information of the system. This is why black holes are holographic.  
 
   The source of the energy gap are quantum fluctuations.  
   The crucial point is that in a quantum theory, we must allow the boundary to fluctuate. 
 For example, for a closed brane bubble, the fluctuations are due to degrees of freedom that compose it.  Similarly, for any finite mass  black hole there is at least one model-independent source of horizon fluctuation: The back-reaction from Hawking evaporation. 
  Once the boundary fluctuations are taken into account, they create 
  bilinear (and higher)  terms in the effective St\"uckelberg action. Such terms generate the energy gap in 
  St\"uckelberg qubits.  On very general grounds, the energy gap can be estimated as 
 (\ref{gap!}),   where  $\alpha(R)$ is the characteristic quantum coupling of the fluctuating degrees of freedom, evaluated at the scale $R$, and 
  $m$ is their characteristic energy scale. In a massless theory it is  given by the inverse size of the system, $m=\hbar/R$.  Or, if the theory has  an intrinsic  mass gap in the spectrum, 
  $m$ is set by the latter.

 In order to see more formally why the gap is generated,  let us rewrite the action (\ref{actionstuck}) in a four-dimensional bulk language 
    \begin{equation} 
 {\mathcal S}_{\eta} \, = \int d^4x \,  \eta_{\mu}(x) \,J^{\mu}(x)\,,  
 \label{coupling}
 \end{equation} 
  where,  $J^{\mu}$ is the Hodge-dual,  
  $J^{\mu}   \equiv \, \epsilon^{\mu\nu\alpha\beta} J_{\nu\alpha\beta}$ \,,
  of the boundary current, 
  \begin{equation}
  J^{\nu\alpha\beta}(x) \equiv  \int dX^{\nu}\wedge dX^{\alpha}\wedge dX^{\beta}  \, \delta^4(x^{\mu} - X^{\mu}) \, .  
  \label{current} 
 \end{equation}   
  Due to virtual quantum processes the current-current correlator is non-vanishing (once interactions are taken into account) and induces a bilinear term in the action, 
 \begin{equation} 
  \eta_{\mu}\eta_{\nu} \langle J^{\mu} J^{\nu} \rangle \,. 
 \label{term}
 \end{equation} 

 Of course, the bilinear term will be generated for the full gauge-invariant combination,  
 \begin{equation}  
  \left( L(h) \, + \, \partial^{\mu} \eta_{\mu} \right)^2 \,.    
  \label{coupling} 
 \end{equation}   
  Thus,  as it is usual,  the St\"uckelberg field acquires an energy gap via a Higgs-type mechanism. The  
 coefficient of this term can only vanish in the limit in which the fluctuations of the boundary can be ignored. 
    The higher is the ability of the boundary to fluctuate, the bigger is the generated mass gap. 
   It is now clear what limits the capacity of information-storage (\ref{Inform}). It is the energy gap, $\Delta$, in the 
   St\"uckelberg qubit.   If $\Delta$ is large, the storage of information in qubits is very costly, and the corresponding 
   system is far from being holographic.  This is the reason why most of the systems with information 
   boundaries are not holographic.  
   
    We are now ready to understand, at least qualitatively,  why the black hole horizon is special among all possible information boundaries.  We shall show this using two different arguments. 
    
   The first argument comes from a straightforward application of (\ref{gap!}) to the black hole case. 
  If we assume that the quantum fluctuations are due to gravitons of wavelength $R$, we have to take for $\alpha(R)$ the expression (\ref{gravitycoupling}) and $m=\hbar/R$.  Substituting both into (\ref{gap!}), we get 
  \begin{equation}
  \Delta_{BH} \, = \, {\hbar L_P^2 \over R^3} \,.   
  \label{deltablack}
  \end{equation}
Remarkably,  as already discussed above,  this gives an information measure (\ref{Inform})  which, as shown at the end of section I,  reproduces 
  Bekenstein entropy.
  
    An alternative argument  that leads us to the same result is based on estimating the strength of quantum fluctuations using the measure of back-reaction from the Hawking radiation.         
    For this,  first recall that the quantum fluctuations of the horizon vanish in the semi-classical limit,
  in which  $L_P \rightarrow 0$ and the black hole size $R$ is kept finite.      
  In this limit the black hole horizon becomes {\it infinitely rigid}  and is insensitive to any type of quantum back-reaction.  In the case of finite $L_P$, the back-reaction on the horizon from Hawking radiation   can easily be estimated to be of order $L_P^2/R^2$, by taking into account the relative change of the black hole mass, $\delta M_{BH}$,  or its temperature,
   due to each Hawking emission, 
   \begin{equation}
   {\delta M_{BH} \over M_{BH}} \, = \,- \, \alpha(R)  \, .    
  \label{backreaction}
   \end{equation} 
    As it is clear, this back-reaction parameter coincides with the coupling of gravitons of wavelength $R$ discussed above, so we have denoted both by the 
   same symbol $\alpha(R)$.   
      Thus, the parameter $\alpha(R)$ is the measure of the black hole quantumness.  It is 
    natural to expect that the quantum-mechanically generated energy gap in St\"uckelberg qubits must be 
    suppressed by this additional factor relative to a gap expected for an ordinary quantum system of size $R$, which would be 
    $\hbar /R$.   This gives a simple alternative  estimate for the quantum-generated energy gap of the St\"uckelberg qubit,    
  \begin{equation}
  \Delta_{BH}  =  \alpha(R) { \hbar \over R} \, , 
  \label{gapalpha}
  \end{equation}  
  which agrees with  (\ref{deltablack}).   
    
    Hence, both estimates give us the same energy gap of St\"uckelberg modes, which leads to the expressions (\ref{Imzero}) and (\ref{areaI}) 
for the  quantity  ${\mathcal I} $ that matches the Bekenstein entropy of a black hole of  radius  $R$.  

   From (\ref{gap!}) it is easy to understand why there  are no systems of size $R$  with 
   the energy gap in the St\"uckelberg fields smaller than the black hole horizon.  Indeed, no matter how weakly the quantum constituents are interacting by other forces, they must interact gravitationally.  Correspondingly, putting aside accidental cancellations, the gravitational coupling $\alpha$ sets the lowest bound on the quantum interaction strength and subsequently on the 
   generated energy gap \footnote{In this light, it would be interesting to see if  the energy gap can be maintained smaller due to cancellations, e.g., by supersymmetry.}. 
   
       For a larger energy gap $\Delta$, the entropy of the bounded inner system  would be much less than the black hole entropy. This allows us to understand the high capacity of black hole information-storage 
    in terms of the lowest mass gap in St\"uckelberg fields.

 \section{Examples with Large Energy Gap for the Boundary St\"uckelberg  Mode} 
  
  In order to get a broader view,  let us now consider other gauge systems, with dynamically-generated information boundaries.   As we shall see, the information measure ${\mathcal I}$ in these systems is very small, due to the energy gap which goes hand in hand with the generation of the boundary.  We shall consider two examples, which in a certain sense are 
  dual to each other.

\subsection{Classical energy gap: Information boundary from Meissner effect} 

 The first example is given by a  slightly-deformed Higgs model that in addition to  the usual Higgs vacuum, with a massive 
 photon, admits a second vacuum in which the photon is massless and the $U(1)$-gauge theory is in the Coulomb
 phase.  
   The simple model is given by the following Lagrangian, 
    \begin{equation}
   L_{\Phi} \, = \, |D_{\mu} \Phi|^2 \, -  \, F_{\mu\nu} F^{\mu\nu} \,  -  \, \lambda^2 {|\Phi|^2 \over v^2} \left (|\Phi|^2 - v^2 \right )^2 \,, 
  \label{Higgs}
  \end{equation} 
  where  $\Phi (x) = |\Phi (x)| e^{i\theta(x)}$ is a complex scalar field,  $F_{\mu\nu} \equiv \partial_{\mu}A_{\nu} - \partial_{\nu}A_{\mu}$ is the Maxwellian field strength and 
  $D_{\mu}  \equiv \partial_{\mu} - ig A_{\mu}$ with $g$ being the gauge coupling.
  The two physically inequivalent vacua are, $\langle |\Phi| \rangle \, = \, 0$ and $\langle |\Phi| \rangle \, = \, v$. 
  In the first vacuum, the propagating fields are the  massless photon and a complex scalar of mass 
  $m_{\Phi} \, = \, \hbar \lambda v$, whereas in the second one we have a massive photon with the mass 
  $m_A = \hbar gv$ and a real Higgs scalar with the mass  $m_{\Phi} =  \hbar \lambda v$.  In both vacua the total number of propagating degrees of freedom is equal to four.  
   
      This theory also admits configurations in which the two vacua coexist and are separated 
  by a domain wall across which the Higgs expectation value interpolates from $0$ to $v$.  Without loss 
  of generality we can place the wall at $z=0$. 
   The thickness of the wall 
  is set by the Compton wavelength of the Higgs particle $ \delta = (\lambda v)^{-1}$.    At energies $\ll m_{\Phi}$ we can  integrate out the thickness of the wall and the effective theory of the photon field becomes, 
 \begin{equation}  
 L_{eff}  =   -  F_{\mu\nu} F^{\mu\nu}  + \hbar^{-2}m^2(z) \left(A_{\mu} - {1 \over g} \partial_{\mu} \theta\right )^2\, ,  
 \label{photoneff} 
 \end{equation} 
  where the function $m^2(z)$ can be approximated as a step function  $m^2(z < 0 ) = 0,~~ m^2(z > 0 ) = m_A$.
 At energies $\ll m_A$ the photon cannot penetrate the Higgs region and the effective theory of the photon 
 is a Coulomb domain bounded by the wall. The eaten-up Goldstone $\theta$ plays the role  
 of the St\"uckelberg field that is maintaining the gauge redundancy,  $A_{\mu} \rightarrow  A_{\mu} + {1 \over g} \partial_{\mu} \omega, ~~~ \theta \rightarrow  \theta + \omega$, despite the presence of the boundary.  
However, there is a mass gap present already at the classical level. In fact, this allows an observer to conclude that the other side of the wall is in the Higgs phase.
   
  Let us take for a region $B$ the interior of a Higgs vacuum bubble of radius $R$, embedded into 
   the Coulomb vacuum (region $A$)\footnote{If the two vacua are exactly degenerate, as it is the case for the Lagrangian
   (\ref{Higgs}),  the finite size bubble cannot be static and will tend to collapse because of the tension of the bubble wall. This is not a big complication for out purposes, if we are interested in information-storage properties  
 over short time intervals. Moreover, we can make the bubble of a given radius static by balancing the tension force 
 by a pressure-difference created via a small difference between the energy densities of the two vacua.  An alternative 
 way is to stabilize the bubble by introducing some charged particles that are localized within the bubble wall. }. Then the region $B$, which is in the Higgs phase, at low energies is  bounded by the sphere.  The St\"uckelberg field is the eaten-up Goldstone that maintains the gauge redundancy throughout the space.  But there is a large energy gap.  The information measure $I$ is always very small, unless  $m_A < \hbar /R$. The latter choice however 
     would mean that effectively there is no information boundary for the photon.  In such a regime, we need to 
     consider the information stored in the Goldstone mode $\theta$ separately.     

   The latter statement becomes obvious by taking an extreme case when we  switch off the gauge coupling completely
 $g =0$, but keep $v$ finite. Then there is a massless Goldstone mode localized within the bubble and the information can be stored in 
 its zero momentum mode.  However,  in this limit the bubble is transparent for the photon and no information boundary exists for the gauge field.

   \subsection{Quantum gap: Boundary from dual Meissner effect} 
   
    Let us consider a situation in which  the information boundary is dynamically  formed by a magnetic monopole condensate. This can be achieved  by employing the massless gauge field localization mechanism of \cite{giamisha},  which is based on interpolation between the confining and Coulomb phases. 
    
    We can achieve this situation by modification of the 
    model (\ref{Higgs}) by embedding the $U(1)$ into a gauge $SU(2)$-symmetry and promoting the Higgs field into a real triplet 
  representation $\Phi^a,~a=1,2,3$.  The Lagrangian now becomes, 
    \begin{equation}
   L_{SU(2)} \, = \, |D_{\mu} \Phi^a|^2 \, -  \, F_{\mu\nu}^a F^{a\mu\nu} \,  -  \, \lambda^2 {\Phi^a\Phi^a \over v^2} \left (\Phi^b\Phi^b - v^2 \right )^2 \, , 
  \label{Higgstripled}
  \end{equation} 
  where $D_{\mu}$ is the usual $SU(2)$-covariant derivative and   $F_{\mu\nu}^a$ is a non-abelian 
  field strength.  We have chosen the Higgs potential as in \cite{monopole1}.  
  Now, the two degenerate vacua are $\langle \Phi^a \rangle =0$ and $\langle \Phi^a \rangle  = \delta^a_3 v$. In the first one, the theory is in the 
  $SU(2)$-phase.   
   In the second one, the $SU(2)$-group is Higgsed down to a $U(1)$ subgroup, which is in the  Coulomb phase. 
  Again,  we can consider a configuration in which the two phases are separated by a domain wall placed 
  at $z=0$.  Notice, unlike the previous example, here at the  classical level the wall is transparent for the $U(1)$-photon.   
  However, in the quantum theory the story changes dramatically.  
  The $SU(2)$-phase becomes confining and develops a mass gap given by the QCD scale  $\Lambda$.
   The photon traveling across the wall can only penetrate the $SU(2)$-domain in form of a glueball
   of mass $\Lambda$.  Correspondingly, at energies below $\Lambda$ the effective theory of 
   the $U(1)$-photon is a gauge theory with a boundary. The gauge invariance is again maintained by a St\"uckelberg field, the role of which is played by the phase of the monopole condensate that gives mass to a magnetic photon (see \cite{monopole1,monopole} for detailed discussion of different aspects of such a picture).   
 But, the very same physics that generates the monopole condensate also generates the mass gap due to the confinement of  the electric  charges. 
   
 In order to make  the example more transparent let us place Alice in the $SU(2)$-domain.  
Imagine that Alice is observing the bubble of the $U(1)$-vacuum.  Let us take the size of the bubble $R$ to be much larger than the QCD length, $R \gg \hbar/\Lambda$.   
 Inside this bubble information can be stored 
in form of the low-energy photon quanta.  If these quanta have energies $\ll \Lambda$, the information stored in them cannot penetrate into the $SU(2)$ -domain, where the energy gap is 
$\Lambda$ \footnote{Recently, a coherent state picture of photons in a ball was studied in \cite{ball}. 
It would be interesting to see if the $SU(2)$-confining region discussed here  provides the boundary conditions imposed in this analysis and also to see if the critical transition can be modeled in the many body photon language, when the ball reaches the size of $SU(2)$ QCD-length, $R \sim \hbar/\Lambda$.}. 

 Thus, the domain wall separating the $U(1)$-Coulomb and  $SU(2)$-confining 
phases acts as an information boundary for Alice.  However, the barrier in this case is 
rather peculiar.  Indeed, the $U(1)$ factor is nowhere Higgsed throughout the space and consequently
there is no potential energy barrier for the photon at the level of the fundamental Lagrangian of the form (\ref{photoneff}).  What prevents the photon to escape out of the bubble is not 
a charge condensate, but rather a condensate of magnetic monopoles. 
 However, the magnetic monopole condensate, unlike the charge condensate,  is {\it not}  generating any fundamental mass-term  for a photon,  but 
 rather  is making it {\it non-propagating}.   
   The mass gap generated for the St\"uckelberg field satisfies the general relation  of the type (\ref{Im}), with the only caveat that the role of $m$ is played by $\Lambda$ and 
   the role of $\alpha(m)$ is played by the QCD coupling $\alpha_{SU(2)}(\Lambda)$ evaluated 
   at the scale $\Lambda$, which is of order one.    
   
     Thus, in both considered examples the mass gap in the St\"uckelberg field is generated simultaneously with the 
  generation  of  the boundary.  In both cases the information boundary is bounding the regions of condensed charges, which are 
  either electric or magnetic. In both cases,  the St\"uckelberg field is the phase of the condensate and in both cases 
  the generation of the mass gap in this degree of freedom is inseparable from the generation of the information boundary 
  \footnote{In all these examples for information measure ${\mathcal  I}$  we get (\ref{Im}) 
  where $m$ is the mass gap of Alice's physics. Obviously this mass gap must be bigger than $\frac{\hbar}{R}$ to allow Alice to know the actual existence of a hidden region of size $R$, so
 in these cases ${\mathcal  I}$ is trivially bounded by (\ref{Imzero}).}.   
  
   The difference is that in the electric condensate case, this phenomenon can take place already at the classical level, whereas in the case of magnetic charges the entire effect  is quantum.

 \section{Relation between Holography and Quantum Criticality} 
 
  According to (\ref{gapmzero}), in massless gauge systems with an information boundary the energy 
  gap in St\"uckelberg qubits is suppressed by the coupling $\alpha$, relative to the typical 
  energy gap, which would be expected to be $\sim \hbar /R$.  This expression suggest an interesting underlying relation with
  quantum criticality.   According to the studies of \cite{critical,gold}, systems at the quantum critical point  are very efficient storers of information, due to the appearance of  qubits  with an energy  
  gap that exhibits a suppression very similar to (\ref{gapmzero}).  In particular, this is a property 
  of a gas of  bosons with attractive interaction strength $\alpha$.  In this case, the criticality is reached when the occupation number is equal to $\alpha^{-1}$.  This system has been studied in a series of papers \cite{critical,gold} from a quantum information perspective, as a prototype model for black holes.   By now it is well established that the quantum critical point  is populated by 
  nearly-gapless qubits. 
   These qubits can be described in different languages. In the many-body language they can be described as Bogoliubov modes of the critical Bose-gas. Alternatively, they can be characterized 
 as the Goldstone modes of a non-linearly realized symmetry of the condensate \cite{gold}. 
 Based on this connection, the black hole portrait of  \cite{us, critical, gold} suggests that black hole information is carried by the Bogoliubov-Goldstone qubits of the critical graviton condensate, with the energy gap
 given by, 
 \begin{equation}
 \Delta_\text{Bogoliubov} \, = \, \alpha(R) {\hbar \over R} \, .
 \label{gapBH}
 \end{equation} 
  The approach developed in the present paper suggests yet another description of these modes, 
 which naturally leads us to a potential link between quantum criticality and holography.  Namely, the striking similarity between the energy gaps  (\ref{gapmzero}) and (\ref{gapBH}) for the St\"uckelberg modes in a gauge description and Bogoliubov
modes in a many-body description of black holes, suggests that we are dealing with the two sides of the same coin and that the holographic systems microscopically represent the systems at a quantum critical point.  The modes that by Alice are seen as  St\"uckelberg degrees of freedom, are, in the microscopic 
many-body description of the black hole, represented as the Bogoliubov-Goldstone  modes of the critical graviton condensate.    It is tempting to generalize this correspondence  to other types 
of holographic systems.  This relation can also work in the opposite direction suggesting that 
critical systems must exhibit some sort of holography in the sense presented here.
In this respect would interesting to investigate the observation of equivalence, in large $N$-limit,  of 
the grounds states of one dimensional $N$-particle Bose-gas on a ring and a gauge  theory on a two-sphere \cite{andredaniel}. 

The connection between the St\"uckelberg formulation of holography and quantum criticality gives an exciting possibility of experimentally studying yet another black hole property in critical systems that can be manufactured in table-top labs, 
in the spirit of \cite{critical, gold}.

    \section{Conclusions} 
    
    In this note we have suggested that some key features of holography can be understood in terms of the basic principles of gauge redundancy.  The maintenance of this redundancy requires the appearance of  St\"uckelberg degrees of freedom at the information boundary.  These modes 
  then act as qubits that store a fraction or the entire information of the inaccessible region.   
 The amount of information stored by St\"uckelberg qubits  is determined by their energy gap
 $\Delta$. 
 Even if absent classically, this gap is generated by quantum effects and is given by 
 (\ref{gap!}).  
 Correspondingly, we have introduced a measure of information capacity ${\mathcal I}$ given by (\ref{Im}), which quantifies how energetically cheap is the excitation of  St\"uckelberg qubits relative to the usual energy gap of a system of the same size, typically expected to be $\hbar/R$. For gauge theories without an intrinsic 
 mass gap in the particle spectrum, the quantities $\Delta$ and  ${\mathcal I}$ are given by 
 (\ref{gapmzero}) and (\ref{Imzero}) respectively,
   \begin{equation}
 \Delta_\text{St\"uckelberg} \, = \, \alpha(R) {\hbar \over R}\,, ~~~
      {\mathcal  I}  =  {1 \over \alpha (R)} \, , 
     \label{Ianddelta}
   \end{equation}
  which we have copied here for the reader's convenience.  For the gravitational coupling $\alpha(R)$, 
  which scales as (\ref{gravitycoupling}), the above expression for ${\mathcal  I}$ 
    coincides with the Bekenstein entropy. 
  The  expression (\ref{Ianddelta}) makes it clear, why among all possible information boundaries
  the black hole horizon can store a maximal amount of information. The reason is simply  that gravity 
  sets the lower bound on how weakly the quantum modes can interact. Even if particles in the theory
  interact via additional hypothetical forces much weaker than gravity, they must still couple gravitationally. 
   Correspondingly, black holes house  St\"uckelberg modes with the smallest possible gap in nature. 
  In this respect, it would be interesting to study what happens for extremal black holes or 
  other systems (e.g., supersymmetric ones) in which gravitational attraction of the constituents can be compensated  
  by the repulsion due to other forces. One may expect that the St\"uckelbergs  for such systems remain gapless.
    
     We thus conclude that black holes carry St\"uckelberg hair, which is closely analogous to  the 
 Bogoliubov-Goldstone hair of the quantum critical picture of \cite{us,critical,hair,gold}.  Both types of hair become infinite but unresolvable in the classical limit, because in this limit $\alpha(R)$ vanishes and 
 ${\mathcal I}$ diverges.

   Finally, taking into account the striking similarities in suppressions of their energy gaps, we have suggested a connection between the St\"uckelbergs  and Bogoliubov modes of the critical graviton condensate.  We have proposed that holographic systems are secretly systems at a quantum critical point. Generalizing this hypothesis in the opposite direction, the usual critical systems that can be obtained  in table top labs, such as critical Bose-Einstein condensates of cold atoms, must also exhibit some notion of holography, which can open up very interesting experimental prospects.

  \section*{Acknowledgements}
The work of G.D. was supported in part by Humboldt Foundation under Humboldt Professorship, 
ERC Advanced Grant 339169 "Selfcompletion'', by TRR 33 "The Dark
Universe" and by the DFG cluster of excellence "Origin and Structure of the Universe". 
The work of C.G. was supported in part by Humboldt Foundation and by Grants: FPA 2009-07908, CPAN (CSD2007-00042) and by the ERC Advanced Grant 339169 "Selfcompletion''.
The work of N.W. was supported by
the Swedish Research Council (VR) through the Oskar Klein Centre.

\end{document}